\documentclass[twocolumn,A4]{article}
\usepackage[dvips]{graphics}
\begin{document}
\onecolumn
\begin{center}

{\bf{\Large Quantum transport through molecular wires}}\\
~\\
Santanu K. Maiti$^{1,2,*}$ and S. N. Karmakar$^1$ \\
~\\
{\em $^1$Theoretical Condensed Matter Physics Division,
Saha Institute of Nuclear Physics, \\
1/AF, Bidhannagar, Kolkata-700 064, India \\
$^2$Department of Physics, Narasinha Dutt College,
129, Belilious Road, Howrah-711 101, India} \\
~\\
{\bf Abstract}
\end{center}
We explore electron transport properties in molecular wires made
of heterocyclic molecules (pyrrole, furan and thiophene) by using the
Green's function technique. Parametric calculations are given based on
the tight-binding model to describe the electron transport in these 
wires. It is observed that the transport properties are significantly 
influenced by (a) the heteroatoms in the heterocyclic molecules and 
(b) the molecule-to-electrodes coupling strength. Conductance ($g$) 
shows sharp resonance peaks associated with the molecular energy levels 
in the limit of weak molecular coupling, while they get broadened in 
the strong molecular coupling limit. These resonances get shifted with 
the change of the heteroatoms in these heterocyclic molecules. All the 
essential features of the electron transfer through these molecular 
wires become much more clearly visible from the study of our 
current-voltage ($I$-$V$) characteristics, and they provide several 
key informations in the study of molecular transport.
\vskip 1cm
\begin{flushleft}
\end{flushleft}
\vskip 4in
\noindent
{\bf ~$^*$Corresponding Author}: Santanu K. Maiti

Electronic mail: santanu.maiti@saha.ac.in
\newpage
\twocolumn

\section{Introduction}

Quantum transport properties of organic molecules bridging over electrodes
are recent interest of nanotechnologies since they constitute promising
building blocks for future generation of nanoelectronic devices where the
electron transport is predominantly coherent.\cite{nitzan1,nitzan2}
Following experimental developments, theory can play a major role
in understanding the new mechanisms of conductance, and in the last few 
decades, electron transport through different nano-scale 
systems\cite{orella1,orella2,cron,holl} have been studied enormously. 
The single-molecule electronics plays a key role in the design of future 
nanoelectronic circuits, but the goal of developing a reliable 
molecular-electronics technology is still over the horizon and many key 
problems, such as device stability, reproducibility and the control of 
single-molecule transport need to be solved. Starting with the paper of 
Aviram and Ratner\cite{aviram} in which a molecular electronic device 
has been suggested for the first time, the development of a theoretical 
description of molecular electronic devices has been pursued. Since then 
several numerous experiments\cite{metz,fish,reed1,reed2,tali} have been 
performed through molecules placed between two metallic electrodes with 
few nanometer separation. The operation of such two-terminal devices 
is due to an applied bias. Current passing across the junction is 
strongly nonlinear function of the applied bias voltage and its detailed 
description is a very complex problem. The complete knowledge of the
conduction mechanism in this scale is not well understood even today.
For example, it is not very clear how the molecular transport is affected 
by the structure of the molecule itself or by the nature of its coupling 
with the electrodes. The most important issue is probably the quantum 
interference effects\cite{baer2,baer3,tagami,walc1} among the electron 
waves traversing through different arms of the molecule. Another important 
issue is the molecular coupling to the side attached electrodes.\cite{baer1}
Tuning this coupling, one can control the current amplitude across the 
bridge quite significantly. Similar to these, there are several other 
factors like the electron-electron correlations,\cite{tom} dynamical 
fluctuations,\cite{blanter,walc2} etc., which provide rich effects in the 
electron transport. To design molecular electronic devices with specific 
properties, structure-conductance relationships are also needed, and in 
a very recent work Ernzerhof {\em et al.}\cite{ern1} have presented a 
general design principle and performed several model calculations to 
demonstrate the concept. Here we focus on single molecular transport 
that are currently the subject of substantial experimental, theoretical 
and technological interest. These molecular systems can act as gates, 
switches, or transport elements, providing new molecular functions that 
need to be well characterized and understood. In many molecular devices, 
electron transport is dominated by conduction through broadened HOMO or 
LUMO states. In contrast, in this article we find that the electron transport 
through molecular bridges can be controlled very sensitively by chemically 
modifying the heteroatoms in the heterocyclic molecules.

Here we demonstrate that there are advantages in engineering molecules which
can be modified externally to achieve control over transport by altering the
properties of the heteroatoms. In an efficient molecular transport system,
actual contact of the molecule to both electrodes is required. Therefore the
simplest theoretical view is based on the tight-binding type one-electron
picture. In this article we reproduce an analytic approach based on the
tight-binding model to investigate the electron transport properties for the
model of single heterocyclic molecules named as pyrrole, furan and thiophene, 
and the coupling of the molecules to the side attached electrodes is treated 
through the Newns-Anderson chemisorption theory.\cite{new,muj1,muj2} There 
exist several {\em ab initio} methods for the calculation of the 
conductance\cite{yal,ven,xue,tay,der,dam,ern3,zhu1,zhu2,cheng1,cheng2} 
as well as model calculations.\cite{muj1,muj2,sam,hjo} The model 
calculations are motivated by the fact that the {\em ab initio} theories 
are computationally too expensive, while the model calculations by using 
the tight-binding formulation are computationally very cheap and also 
provide a worth insight to the problem. In our present study, attention is 
drawn on the qualitative behavior of the physical quantities rather than 
the quantitative ones. Here we do not take into account the effect of 
charge transfer from the electrode which may play a significant role 
in the {\em ab initio} study.

Our scheme of study is as follows. In Section $2$ we describe very briefly
about the methodology for the calculation of the transmission probability
($T$) and the current ($I$) through a finite size conducting system
sandwiched between two metallic electrodes by the use of Green's function
technique. Section $3$ investigates the behavior of the conductance ($g$)
as a function of the injecting electron energy ($E$) and the current-voltage
($I$-$V$) characteristics for the model of three different heterocyclic
molecules. Here we focus our results in the aspects of (a) the heteroatoms
in the heterocyclic molecules, and (b) the molecular coupling to the side
attached electrodes. Finally, we summarize our results in Section $4$.

\section{A Brief Description Onto the Theoretical Formulation}

In this section we discuss very briefly about the methodology for the
calculation of the transmission probability ($T$), conductance ($g$) and
current ($I$) through a finite size conductor attached to two 
one-dimensional semi-infinite metallic electrodes by the use of Green's
function technique.

\begin{figure}[ht]
{\centering \resizebox*{7.5cm}{1.35cm}{\includegraphics{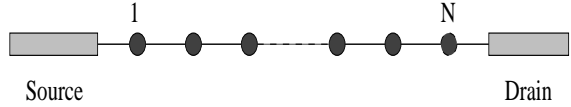}}\par}
\caption{Schematic view of a one-dimensional conductor with $N$ number of
atomic sites (filled circles) attached to two metallic electrodes (source
and drain) through the atomic sites $1$ and $N$ respectively.}
\label{dot}
\end{figure}

Let us refer to Fig.~\ref{dot}, where a one-dimensional conductor with 
$N$ number of atomic sites (array of filled circles) connected to two 
semi-infinite metallic electrodes,
viz, source and drain. The conducting system within the two electrodes 
can be anything like a single molecule, or an array of few molecules, or 
an array of some quantum dots, etc. At sufficient low temperature and 
applied bias voltage, the conductance of the conductor can be written by
the Landauer conductance formula\cite{datta,marc} as,
\begin{equation}
g=\frac{2e^2}{h}T
\label{land}
\end{equation}
where $g$ is the conductance and $T$ is the transmission probability of an
electron through the conductor. The transmission probability can be expressed
in terms of the Green's function of the conductor and its coupling to the
side attached electrodes by the relation,\cite{datta,marc}
\begin{equation}
T=Tr\left[\Gamma_S G_c^r \Gamma_D G_c^a\right]
\label{trans1}
\end{equation}
where $G_c^r$ and $G_c^a$ are respectively the retarded and advanced Green's
function of the conductor. $\Gamma_S$ and $\Gamma_D$ are the coupling terms
due to the coupling of the conductor to the source and drain respectively. 
For the complete system i.e., the conductor with the two electrodes, the 
Green's function is defined as,
\begin{equation}
G=\left(\epsilon-H\right)^{-1}
\end{equation}
where $\epsilon=E+i\eta$. $E$ is the energy of the source electron and 
$\eta$ gives an infinitesimal imaginary part to $\epsilon$. Evaluation
of this Green's function requires the inversion of an infinite matrix as the
system consists of the finite conductor and the two semi-infinite electrodes.
However, the entire system can be partitioned into sub-matrices corresponding
to the individual sub-systems, and the Green's function for the conductor
can be effectively written as,
\begin{equation}
G_c=\left(\epsilon-H_c-\Sigma_S-\Sigma_D\right)^{-1}
\label{grc}
\end{equation}
where $H_c$ is the Hamiltonian of the conductor sandwiched between the two
electrodes. The Hamiltonian of the conductor in the tight-binding framework
can be written within the non-interacting picture in this form,
\begin{equation}
H_c=\sum_i \epsilon_i c_i^{\dagger} c_i + \sum_{<ij>}t
\left(c_i^{\dagger}c_j + c_j^{\dagger}c_i \right)
\label{hamil1}
\end{equation}
where $c_i^{\dagger}$ ($c_i$) is the creation (annihilation) operator of an
electron at site $i$, $\epsilon_i$'s are the site energies and $t$ is the
nearest-neighbor hopping strength. In Eq.~(\ref{grc}),
$\Sigma_S=h_{Sc}^{\dagger}g_S h_{Sc}$ and
$\Sigma_D=h_{Dc} g_D h_{Dc}^{\dagger}$ are the self-energy operators due to
the two electrodes, where $g_S$ and $g_D$ are respectively the Green's
function for the source and drain. $h_{Sc}$ and $h_{Dc}$ are the coupling
matrices and they will be non-zero only for the adjacent points of the
conductor, $1$ and $N$ as shown in Fig.~\ref{dot}, and the electrodes
respectively. The coupling terms $\Gamma_S$ and $\Gamma_D$ for the conductor
can be calculated through the expression,
\begin{equation}
\Gamma_{S(D)}=i\left[\Sigma_{S(D)}^r-\Sigma_{S(D)}^a\right]
\end{equation}
where $\Sigma_{S(D)}^r$ and $\Sigma_{S(D)}^a$ are the retarded and
advanced self-energies respectively and they are conjugate with each
other. Datta {\em et al.}\cite{tian} have shown that the self-energies
can be expressed like as,
\begin{equation}
\Sigma_{S(D)}^r=\Lambda_{S(D)}-i \Delta_{S(D)}
\end{equation}
where $\Lambda_{S(D)}$ are the real parts of the self-energies which
correspond to the shift of the energy eigenvalues of the conductor, and 
the imaginary parts $\Delta_{S(D)}$ of the self-energies represent the
broadening of these energy levels. This broadening is much larger 
than the thermal broadening, and accordingly, we restrict our all
calculations only at absolute zero temperature. Here we adopt the 
Newns-Anderson chemisorption model\cite{new,muj1,muj2} for the 
description of the electrodes and for the interaction of the electrodes 
with the conductor, where the effect of the electrodes is then formally 
incorporated through the self-energies $\Sigma_S$ and $\Sigma_D$. To 
describe the electrodes (in the form of a semi-infinite one-dimensional 
chain), in this present scheme, we use the similar kind of tight-binding 
model Hamiltonian as presented in Eq.~(\ref{hamil1}), which is parametrized 
by the constant on-site potential $\epsilon_0$ and nearest-neighbor hopping 
integral $v$. From the standpoint of the band theory or molecular orbital, 
the coupling between the conductor and the electrodes can be attributed 
to one-electron hopping processes, where the hopping parameters are 
$\tau_S$ and $\tau_D$ respectively. All these are the essential parameters 
for this particular scheme to describe the electron transport phenomena in 
a bridge system. Now by utilizing the Newns-Anderson type model, we can 
express the conductance in terms of the effective conductor properties 
multiplied by the effective state densities involving the coupling. This 
permits us to study directly the conductance as a function of the properties 
of the electronic structure of the conductor within the bridge.

Since the coupling matrices $h_{Sc}$ and $h_{Dc}$ are non-zero only for
the adjacent points in the conductor, $1$ and $N$ as shown in Fig.~\ref{dot},
the transmission probability becomes,
\begin{equation}
T(E,V)=4\Delta_{11}^S(E,V) \Delta_{NN}^D(E,V)|G_{1N}(E,V)|^2
\label{trans2}
\end{equation}
where $\Delta_{11}=<1|\Delta|1>$, $\Delta_{NN}=<N|\Delta|N>$ and
$G_{1N}=<1|G_c|N>$.

The current passing across the conductor is depicted as a single-electron
scattering process between the two reservoirs of charge carriers. The
current-voltage relation is evaluated from the following 
expression,\cite{datta}
\begin{equation}
I(V)=\frac{e}{\pi \hbar}\int_{E_F-eV/2}^{E_F+eV/2} T(E,V) dE
\label{curr1}
\end{equation}
where $E_F$ is the equilibrium Fermi energy. For the sake of simplicity, here
we assume that the entire voltage is dropped across the conductor-electrode
interfaces and this assumption doesn't greatly affect the qualitative aspects
of the $I$-$V$ characteristics. This assumption is based on the fact that the
electric field inside the conductor, especially for short conductors, seems
to have a minimal effect on the conductance-voltage characteristics. On the
other hand for quite larger conductors and high bias voltages, the electric
field inside the conductor may play a more significant role depending on the
internal structure and size of the conductor,\cite{tian} but yet the effect
is too small. Using the expression of $T(E,V)$ as in Eq.~(\ref{trans2}) the
final form of $I(V)$ will be,
\begin{eqnarray}
I(V) &=& \frac{4e}{\pi \hbar}\int_{E_F-eV/2}^{E_F+eV/2}\Delta_{11}^S(E,V)
\Delta_{NN}^D(E,V) \nonumber \\
& & \times |G_{1N}(E,V)|^2 dE
\label{curr}
\end{eqnarray}
Eq.~(\ref{land}), Eq.~(\ref{trans2}) and Eq.~(\ref{curr}) are the final
working formule for the calculation of the conductance $g$, transmission
probability $T$, and current $I$ respectively through any finite size
conductor sandwiched between two metallic electrodes.

In this paper, we will describe the electron transport properties 
by using the above methodology for the three different models of 
heterocyclic molecules those are defined as pyrrole, furan and thiophene
(Fig.~\ref{heterocyclic}). For simplicity, we take the unit $c=e=h=1$ in 
our present calculations.

\section{Results and their Interpretation}

\begin{figure}[ht]
{\centering \resizebox*{7cm}{11cm}{\includegraphics{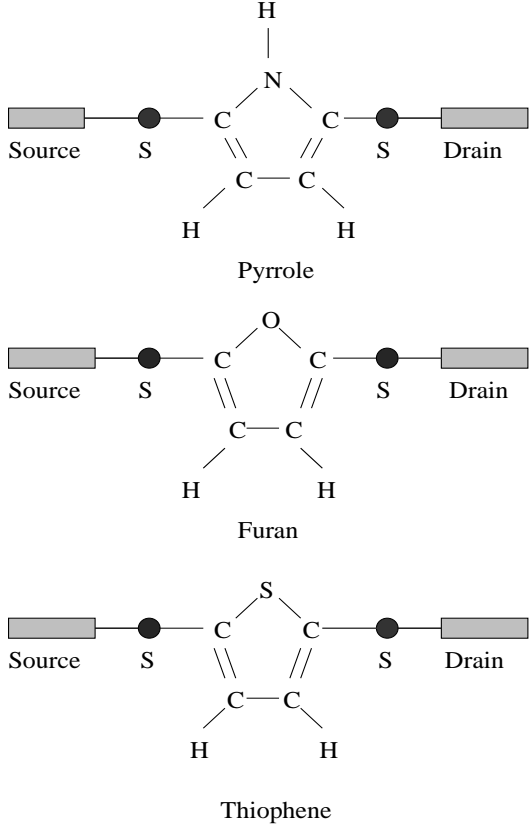}}\par}
\caption{Schematic view of the three different molecular wires. The
heterocyclic molecules (pyrrole, furan and thiophene) are attached to the
electrodes, viz, source and the drain via thiol (S-H) groups.}
\label{heterocyclic}
\end{figure}

This section investigates the behavior of the conductance $g$ as a
function of the injecting electron energy $E$, and the variation of the
current $I$ with the applied bias voltage $V$ for the three different
molecular wires containing the heterocyclic molecules. Here we focus our 
results on the electron transport properties considering the effects of 
(a) the heteroatoms in the heterocyclic molecules (pyrrole, furan and 
thiophene), and (b) the molecule-to-electrodes coupling strength. The 
arrangements of the three different molecular wires are shown in 
Fig.~\ref{heterocyclic}, where nitrogen, oxygen and sulfur are the 
heteroatoms in the molecular wires consisting with pyrrole, furan
and thiophene molecules respectively. These molecules are attached to
the semi-infinite metallic electrodes by thiol (sulfur-hydrogen i.e.,
S-H bond) groups. In actual experimental set-up, the electrodes made
from gold (Au) are used and the molecule coupled to the electrodes
through thiol groups in the chemisorption technique where hydrogen (H)
atoms remove and sulfur (S) atoms reside. To describe these heterocyclic
molecules, we use the similar kind of non-interacting tight-binding 
Hamiltonian as given in Eq.~(\ref{hamil1}). 

All the essential features of this article are studied in the two distinct
regimes. One is so-called the weak-coupling regime, defined by the condition
$\tau_{\{S,D\}} <<t$. The other one is so-called the strong-coupling regime,
denoted by the condition $\tau_{\{S,D\}}\sim t$. For these two limiting
cases, we take the values of the different parameters as follows:
$\tau_S=\tau_D=0.5$; $t=2.5$ (weak coupling) and $\tau_S=\tau_D=2$; $t=2.5$
(strong-coupling). Here we set the on-site energy $\epsilon_0=0$ (we can
take any constant value of it instead of zero since it gives only the
reference energy level) for the electrodes, and the hopping strength $v=4$
in the two semi-infinite metallic electrodes. For the sake of simplicity,
here we set the Fermi energy $E_F=0$.

\begin{figure}[ht]
{\centering \resizebox*{7.5cm}{9cm}{\includegraphics{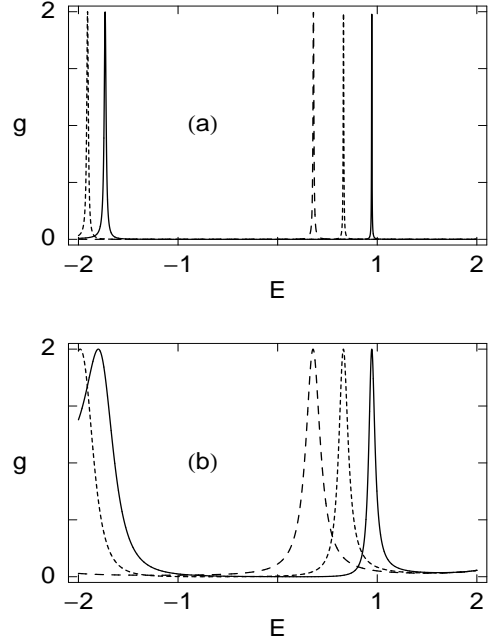}}\par}
\caption{Conductance $g$ as a function of the injecting electron energy
$E$, where the solid, dotted and dashed curves correspond to the molecular 
wires with pyrrole, furan and thiophene molecules respectively.
(a) weak-coupling limit and (b) strong-coupling limit.}
\label{cond}
\end{figure}

In Fig.~\ref{cond}, we plot the conductance $g$ as a function of the injecting
electron energy $E$ for the three molecular wires, where the solid, dotted
and dashed curves correspond to the results for the wires containing
pyrrole, furan and thiophene molecules respectively. The results for the
weak-coupling case are shown in Fig.~\ref{cond}(a), while Fig.~\ref{cond}(b)
gives the results in the limit of strong molecular coupling. In the weak
molecular coupling, the $g$-$E$ characteristics show sharp resonance peaks
for some particular energy values, while they (resonance peaks) disappear
for all other energies. These resonance peaks are associated with the 
energy eigenvalues of the individual molecules. Therefore we can say
that, the conductance spectrum manifests itself the electronic structure 
of the molecule. As expected, the maximum value of these conductance 
peaks goes to two i.e., the transmission probability ($T$) becomes unity 
since we get the relation $g=2T$ from the Landauer conductance formula, 
Eq.~(\ref{land}), with $e=h=1$ in our present formulation. From these 
results it is observed that the resonance peaks get shifted in the scale 
of the energy $E$ as we chemically change the heteroatoms in these 
heterocyclic molecules. Thus we can predict that the on/off state of the 
electron conduction across the
molecule can be tuned by chemically modifying the heteroatoms, though in
all these three molecular wires the molecular structures are the same. This
provides an important finding in the study of molecular transport. In the 
strong molecule-to-electrodes coupling limit, these resonance peaks get 
substantial widths as presented in Fig.~\ref{cond}(b). Such increment of 
the resonance widths is due to the broadening of the molecular energy levels 
in the limit of strong molecular coupling, where the contribution comes 
from the imaginary parts of the self-energies $\Sigma_S$ and 
$\Sigma_D$,\cite{datta} as mentioned earlier in Section $2$. From the curves 
plotted in Figs.~\ref{cond}(a) and (b) it is observed that, the positions 
of the resonance peaks are independent of the molecule-to-electrodes
coupling strength. Another significant feature observed from these curves
is that, in the strong-coupling limit, the molecular bridge remains in the
on-state condition i.e., electron passes through the molecule for a wide
range of energies, while a fine tuning in the energy scale is necessary
to get the on-state condition for that bridge in the limit of weak molecular
coupling. This feature is quite significant in fabrication of efficient
electronic circuits by using these molecules.

\begin{figure}[ht]
{\centering \resizebox*{7.5cm}{9cm}{\includegraphics{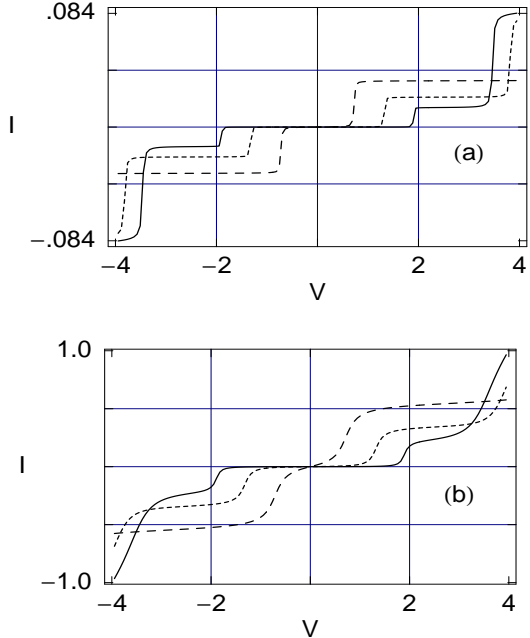}}\par}
\caption{Current $I$ as a function of the applied bias voltage $V$, where
the solid, dotted and dashed curves correspond to the molecular wires
with pyrrole, furan and thiophene molecules respectively. (a) weak-coupling 
limit and (b) strong-coupling limit.}
\label{current}
\end{figure}

The scenario of the electron transfer through such molecular wires becomes
much more clearly visible by describing the current-voltage characteristics,
where the current passing across the molecule is computed from the
integration procedure of the transmission function $T$ (see Eq.~(\ref{curr1})).
The variation of this transmission function looks similar to that of the
conductance spectra (see Figs.~\ref{cond}(a) and (b)), differ only in
magnitude by the factor $2$, since the relation $g=2T$ holds from the
Landauer conductance formula as stated earlier. In Fig.~\ref{current}, we
display the current $I$ as a function of the applied bias voltage $V$ for
the three molecular wires where the solid, dotted and dashed curves
correspond to the same molecular systems as given in Fig.~\ref{cond}. The
results for the weak-coupling limit are shown in Fig.~\ref{current}(a),
while Fig.~\ref{current}(b) corresponds to the results in the limit of
strong molecular coupling. In the limit of weak molecular coupling, the
current-voltage characteristics give staircase-like behavior with sharp
steps (Fig.~\ref{current}(a)). This is due to the existence of the sharp
resonance peaks in the conductance spectrum (Fig.~\ref{cond}(a)) since
the current is evaluated from the integration procedure of the transmission
function $T$. With the increase of the applied bias voltage, the
electrochemical potentials on the electrodes are shifted gradually, and
eventually cross one of the molecular energy levels. Accordingly, a current
channel is opened up and the current-voltage curve produces a jump. The
shape and height of these current steps depend on the width of the molecular
resonances. With the increase of the molecular coupling strength, the
current varies quite continuously with the bias voltage $V$, as illustrated
in Fig.~\ref{current}(b), and achieves much higher values. This can be
understood by noting the areas under the curves in the conductance spectrum
as plotted in Fig.~\ref{cond}(b) in the limit of strong molecular coupling.
From these results we predict that one can achieve much higher current
amplitude in a molecular bridge system just by tuning the
molecule-to-electrodes coupling strength, without changing any geometry of
the bridge. The other most important feature appears from these
current-voltage characteristics is that, the threshold bias voltage ($V_{th}$)
of the electron conduction changes with the change of the heteroatoms in
these heterocyclic molecules. Thus we can tune $V_{th}$ by chemically
modifying the heteroatoms and this provides a key result in the study
of molecular transport in these molecular wires.

\section{Conclusion}

In conclusion, we have explored the electron transport properties of the 
molecular wires made by the heterocyclic molecules (pyrrole, furan and 
thiophene), and observed that the transport properties are significantly 
influenced by (a) the heteroatoms in the molecules and (b) the 
molecule-to-electrodes coupling strength. Here we have used the simple 
tight-binding model to describe the molecular wires, and introduced a 
parametric approach to study the electron transport.

In the study of the $g$-$E$ characteristics we have noticed that the
conductance shows very sharp resonance peaks (Fig.~\ref{cond}(a)) in the
limit of weak molecular coupling associated with the molecular energy
levels, while the widths of these resonances get enhanced substantially
(Fig.~\ref{cond}(b)) in the strong molecular coupling limit. This is due
to the broadening of the molecular energy levels in the limit of strong
coupling, where the contribution comes from the imaginary parts of the two
self energies $\Sigma_S$ and $\Sigma_D$.\cite{datta}

Next we have concentrated our study on the current-voltage characteristics
from which the scenario of the electron transfer through the molecules can
be understood much more clearly. The current changes its behavior from the
staircase-like structure with sharp steps (Fig.~\ref{current}(a)) to the
continuous one (Fig.~\ref{current}(b)) as the molecular coupling changes
its strength from the weak regime to the strong one. From this study we
have observed that the threshold bias voltage $V_{th}$ of the electron
conduction can be tuned by chemically modifying the heteroatoms in these
heterocyclic molecules.

Throughout our study we have used several realistic assumptions. More studies
are expected to take the Schottky effect, comes from the charge transfer
across the metal-molecule interfaces, the static Stark effect, which is
taken into account for the modification of the electronic structure of the
molecular bridge due to the applied bias voltage (essential especially for
higher voltages). However all these effects can be included into our framework
by a simple generalization of the given formalism described here. In this
article we have also ignored the effects of all inelastic scattering
processes, by assuming that the electrons move smoothly from the source to
the drain subject only to elastic scattering within the junction, and
electron-electron correlation to characterize the electronic transport
through such heterocyclic molecules.


\begin{thebibliography}{99}

\bibitem{nitzan1} A. Nitzan, Annu. Rev. Phys. Chem. \textbf{52}, 681 (2001).
\bibitem{nitzan2} A. Nitzan and M. A. Ratner, Science \textbf{300}, 1384
(2003).
\bibitem{orella1} P. A. Orellana, M. L. Ladron de Guevara, M. Pacheco
and A. Latge, Phys. Rev. B \textbf{68}, 195321 (2003).
\bibitem{orella2} P. A. Orellana, F. Dominguez-Adame, I. Gomez and
M. L. Ladron de Guevara, Phys. Rev. B \textbf{67}, 085321 (2003).
\bibitem{cron} S. M. Cronenwett, T. H. Oosterkamp and L. P. Kouwenhoven,
Science \textbf{281}, 5 (1998).
\bibitem{holl}  A. W. Holleitner, R. H. Blick, A. K. Huttel, K. Eber and
J. P. Kotthaus, Science \textbf{297}, 70 (2002).
\bibitem{aviram} A. Aviram and M. Ratner, Chem. Phys. Lett. \textbf{29},
277 (1974).
\bibitem{metz} R. M. Metzger {\em et al.}, J. Am. Chem. Soc. \textbf{119},
10455 (1997).
\bibitem{fish} C. M. Fischer, M. Burghard, S. Roth and K. V. Klitzing,
Appl. Phys. Lett. \textbf{66}, 3331 (1995).
\bibitem{reed1} J. Chen, M. A. Reed, A. M. Rawlett and J. M. Tour, Science
\textbf{286}, 1550 (1999).
\bibitem{reed2} M. A. Reed, C. Zhou, C. J. Muller, T. P. Burgin and J. M.
Tour, Science \textbf{278}, 252 (1997).
\bibitem{tali} T. Dadosh, Y. Gordin, R. Krahne, I. Khivrich, D. Mahalu, 
V. Frydman, J. Sperling, A. Yacoby and I. Bar-Joseph, Nature \textbf{436},
677 (2005).
\bibitem{baer2} R. Baer and D. Neuhauser, J. Am. Chem. Soc. \textbf{124},
4200 (2002).
\bibitem{baer3} D. Walter, D. Neuhauser and R. Baer, Chem. Phys.
\textbf{299}, 139 (2004).
\bibitem{tagami} K. Tagami, L. Wang and M. Tsukada, Nano Lett. \textbf{4},
209 (2004).
\bibitem{walc1} K. Walczak, Cent. Eur. J. Chem. \textbf{2}, 524 (2004).
\bibitem{baer1} R. Baer and D. Neuhauser, Chem. Phys. \textbf{281}, 353
(2002).
\bibitem{tom} T. Kostyrko and B. R. BuÅa, Phys. Rev. B \textbf{67},
205331 (2003).
\bibitem{blanter} Y. M. Blanter and M. Buttiker, Phys. Rep. \textbf{336},
1 (2000).
\bibitem{walc2} K. Walczak, Phys. Stat. Sol. (b) \textbf{241}, 2555 (2004).
\bibitem{ern1} M. Ernzerhof, M. Zhuang and P. Rocheleau, J. Chem. Phys.
\textbf{123}, 134704 (2005).
\bibitem{new} D. M. Newns, Phys. Rev. \textbf{178}, 1123 (1969).
\bibitem{muj1} V. Mujica, M. Kemp and M. A. Ratner, J. Chem. Phys. 
\textbf{101}, 6849 (1994).
\bibitem{muj2} V. Mujica, M. Kemp, A. E. Roitberg and M. A. Ratner, 
J. Chem. Phys. \textbf{104}, 7296 (1996).
\bibitem{yal} S. N. Yaliraki, A. E. Roitberg, C. Gonzalez, V. Mujica
and M. A. Ratner, J. Chem. Phys. \textbf{111}, 6997 (1999).
\bibitem{ven} M. Di Ventra, S. T. Pantelides and N. D. Lang, Phys. Rev.
Lett. \textbf{84}, 979 (2000).
\bibitem{xue} Y. Xue, S. Datta and M. A. Ratner, J. Chem. Phys. \textbf{115},
4292 (2001).
\bibitem{tay} J. Taylor, H. Gou and J. Wang, Phys. Rev. B \textbf{63},
245407 (2001).
\bibitem{der} P. A. Derosa and J. M. Seminario, J. Phys. Chem. B \textbf{105},
471 (2001).
\bibitem{dam} P. S. Damle, A. W. Ghosh and S. Datta, Phys. Rev. B 
\textbf{64}, R201403 (2001).
\bibitem{ern3} M. Ernzerhof and M. Zhuang, Int. J. Quantum Chem.
\textbf{101}, 557 (2005).
\bibitem{zhu1} M. Zhuang, P. Rocheleau and M. Ernzerhof, J. Chem. Phys.
\textbf{122}, 154705 (2005).
\bibitem{zhu2} M. Zhuang and M. Ernzerhof, Phys. Rev. B \textbf{72},
073104 (2005).
\bibitem{cheng1} W. W. Cheng, H. Chen, R. Note, H. Mizuseki and Y. Kawazoe,
Physica E \textbf{25}, 643 (2005).
\bibitem{cheng2} W. W. Cheng, Y. X. Liao, H. Chen, R. Note, H. Mizuseki 
and Y. Kawazoe, Phys. Lett. A \textbf{326}, 412 (2004).
\bibitem{sam} M. P. Samanta, W. Tian, S. Datta, J. I. Henderson and
C. P. Kubiak, Phys. Rev. B \textbf{53}, R7626 (1996).
\bibitem{hjo} M. Hjort and S. Staftr\"{o}m, Phys. Rev. B \textbf{62}, 
5245 (2000).
\bibitem{datta} S. Datta, {\em Electronic transport in mesoscopic systems},
Cambridge University Press, Cambridge (1997).
\bibitem{marc} M. B. Nardelli, Phys. Rev. B \textbf{60}, 7828 (1999).
\bibitem{tian} W. Tian, S. Datta, S. Hong, R. Reifenberger, J. I. Henderson
and C. I. Kubiak, J. Chem. Phys. \textbf{109}, 2874 (1998).

\end{thebibliography}
\end{document}